# Quantum Control of the Time-Dependent Interaction between a Three-Level Ξ-Type Atom and a Two-Mode Field with Damping Term

*Sameh T. Korashy*[1]

[1] *Department of Mathematics, Faculty of Science, Sohag University, 82524, Sohag, Egypt*

**Abstract** The purpose of this paper is to investigate some properties through a three-level Ξ-type atom interacting with a two-mode field. We test this system in the presence of the photon assisted atomic phase damping, detuning parameter and Kerr nonlinearity. Also, the coupling parameter modulated to be time-dependent. The problem solution of this model is given by using the Schrődinger equation when the atom and the field are initially prepared in the excited state and coherent state, respectively. We used the results to calculate some aspects such as atomic population inversion and concurrence. The results show that the time-dependent coupling parameter and the detuning parameter can be considered as a quantum control parameters of the atomic population inversion and quantum entanglement in the considered model.

## 1  Introduction

One of the most famous models in quantum optics is the Jaynes-Cummings model (JCM) [1], which includes of a single two level atom interacting with a single near-resonant quantized cavity mode of the electromagnetic field. It plays a fundamental rule in quantum optics due to the experimental realization of the nonclassical effects. In the fact that it is experimentally realizable and has undergone many theoretical studies [2].

Several modifications and generalizations has been done on the JCM in many different directions such as multi-photon transition, multi-level atoms, intensity-dependent coupling, multi-atoms interaction, multi-mode fields, Stark shift and Kerr nonlinearity have been studied in recent decades [3-16].

A lot of researches are focused on studying multi-level atomic system in different areas of quantum optics. One of the interesting example of the generalization JCM is the system of three-level atom different configurations (Λ, $V$, and Ξ) and one- or two-mode field [3, 8, 17-21]. Many studies has been done on the atom-field entanglement and geometric phase in such

---

[1]  E-mail address: skorashe@yahoo.com

systems [3, 8, 16, 19-22]. A lot of studies of a three-level atom in motion which interacts with a single-mode field in an optical cavity in an intensity-dependent coupling regime have been studied [23]. Dynamics of entropy and nonclassical properties of the state of $\Lambda$-type three-level atom interacting with a single-mode cavity field with intensity-dependent coupling in a Kerr medium have been studied in [24]

The damping is a well known phenomenon in quantum information processing. Several papers have studied the damping effects on entanglement and some non-classical properties. Several studies of the phase damping in the JCM has been studied [25-27] and its influence in quantum properties of the multi-quanta two-mode JCM has been investigated [28]. The time-dependent interaction between a three-level $\Lambda$-type atom two-mode electromagnetic field in a Kerr-like medium, where the field and the atom are suffering decay rates has been studied [29]. Also, The time-dependent interaction between two- $\Lambda$-type three-level atoms and a single - mode cavity field has been discussed [30] when the damping parameter is taken into account.

In the recent years, much attention has been focused on the properties multi atoms and multi-level atomic systems when time-dependent coupling with the field is considered [29-35]. More recently, the entanglement and entropy squeezing for moving two two-level atoms interaction with a radiation field have been investigated in [36].

In order to discuss the dynamics of the present system, we will find the solution of the wave function in the Schrődinger picture under certain approximation similar to that of the Rotating-Wave Approximation (RWA) at any time $t > 0$. This is performed in the next section where we also derive the reduced density matrix of the atom. In Sec. 3, we employ our results to calculate the atomic population inversion and the dynamical properties for different regimes. We devote Sec. 4 to the discussion of the degree of entanglement, where we use the definition of concurrence. Finally, we give our conclusions in Sec. 5.

## 2 Physical Model

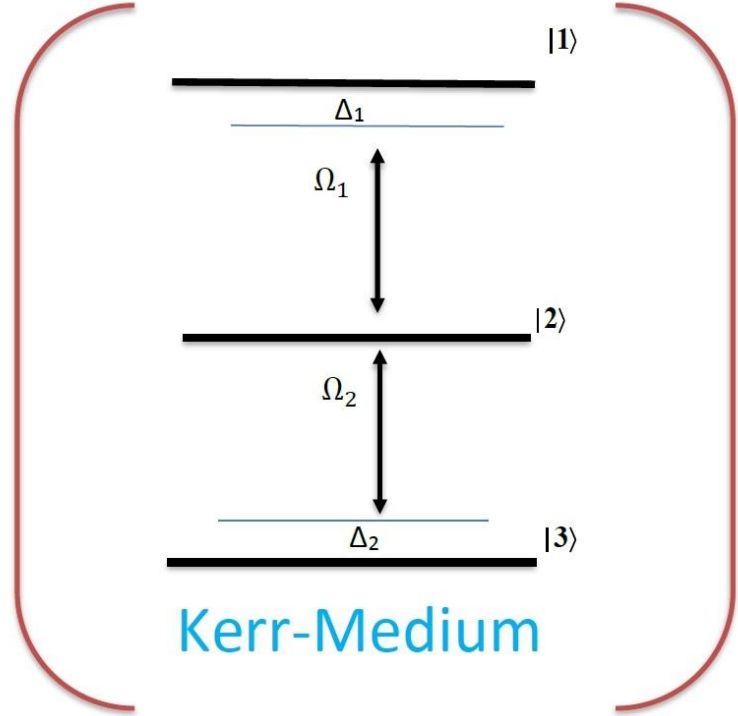

**Fig. 1** Schematic diagram of a three-level Ξ-type atom interacting with a two-mode field.

The considered model is a time-dependent regime consists of a moving three-level (Ξ-type) atom with the energy levels $\omega_1 > \omega_2 > \omega_3$, interacts with a two-mode field of frequency $\Omega_j$ in an optical cavity surrounded by Kerr nonlinearity in the presence of detuning parameters. The transitions $|1\rangle \leftrightarrow |2\rangle$, and $|2\rangle \leftrightarrow |3\rangle$ are allowed while the transition $|1\rangle \leftrightarrow |3\rangle$, is forbidden as shown in Fig. 1. The interaction Hamiltonian in the Rotating-Wave Approximation (RWA) of the introduced physical system [37] ($\hbar = 1$):

$$\widehat{H}_I = f_1(t)(\hat{a}e^{i\Delta_1 t}\hat{\sigma}_{12} + \hat{a}^\dagger e^{-i\Delta_1 t}\hat{\sigma}_{21}) + f_2(t)(\hat{b}\ e^{-i\Delta_2 t}\hat{\sigma}_{23} + \hat{b}^\dagger\ e^{i\Delta_2 t}\hat{\sigma}_{32})$$
$$+\chi_1 \hat{a}^{\dagger 2}\hat{a}^2 + \chi_2\ \hat{b}^{\dagger 2}\ \hat{b}^2 - \frac{i}{2}\gamma_1 \hat{n}_1(\hat{\sigma}_{11} + \hat{\sigma}_{22}) - \frac{i}{2}\gamma_2 \hat{n}_2(\hat{\sigma}_{22} + \hat{\sigma}_{33}), \qquad (1)$$

In which, the operators $\hat{\sigma}_{ii} = |i\rangle\langle j|$ is the atomic raising or lowering operator, the operators $\hat{a}^\dagger(\hat{a})$, are the field creation and annihilation operators of the field mode, respectively, $f_i(t)$, $i = 1,2$, are the atom-field coupling parameters, $\chi_j$ is the third-order nonlinearity of the Kerr-medium and $\gamma_i$, $i = 1,2$, is the photon assisted atomic phase damping parameters which is

positive and real. The detuning parameters $\Delta_1$, $\Delta_2$, are given by
$$\Delta_1 = \omega_1 - \omega_2 - \Omega_1,$$
$$\Delta_2 = \Omega_2 - (\omega_2 - \omega_3), \tag{2}$$
$\Omega_j$, $j = 1,2$ is the frequency of the field mode. We consider $f_1(t) = f_2(t) = f(t) = \lambda_j \cos(\mu t) = \frac{\lambda_j}{2}(e^{i\mu t} + e^{-i\mu t})$, where, $\lambda_j$, $\mu$, $j = 1,2$ are arbitrary constants. As one can see there are two exponential terms in the Hamiltonian: one is rapidly oscillating terms $e^{\pm i(\Delta_j + \mu)t}$ and the other is slowly varying terms $e^{\pm i(\Delta_j - \mu)t}$. In this case if we neglect the rapidly varying terms compared with the slowly varying terms, then the interaction Hamiltonian can be rewritten in the following manner

$$\hat{H}_I = \frac{\lambda_1}{2}(\hat{a}e^{i\delta_1 t}\hat{\sigma}_{12} + \hat{a}^\dagger e^{-i\delta_1 t}\hat{\sigma}_{21}) + \frac{\lambda_2}{2}(\hat{b}e^{-i\delta_2 t}\hat{\sigma}_{23} + \hat{b}^\dagger e^{i\delta_2 t}\hat{\sigma}_{32})$$
$$+\chi_1 \hat{a}^{\dagger 2}\hat{a}^2 + \chi_2 \hat{b}^{\dagger 2}\hat{b}^2 - \frac{i}{2}\gamma_1 \hat{n}_1(\hat{\sigma}_{11} + \hat{\sigma}_{22}) - \frac{i}{2}\gamma_2 \hat{n}_2(\hat{\sigma}_{22} + \hat{\sigma}_{33}) \tag{3}$$

where
$$\delta_1 = \Delta_1 - \mu, \quad \delta_2 = \Delta_2 - \mu. \tag{4}$$

We assume that the wave function of the atom-field at any time $t > 0$ can be expressed as
$$|\Psi(t)\rangle = \sum_{n_1,n_2=0}^{\infty}[G_1(n_1,n_2,t)|1,n_1,n_2\rangle + G_2(n_1+1,n_2,t)|2,n_1+1,n_2\rangle$$
$$+G_3(n_1+1,n_2+1,t)|3,n_1+1,n_2+1\rangle] \tag{5}$$

To reach this goal, suppose that the atom-field initial state is
$$|\Psi(0)\rangle = \sum_{n_1,n_2=0}^{\infty} q_{n_1}q_{n_2}|1,n_1,n_2\rangle \tag{6}$$

where $q_{n_i} = e^{-|\alpha_i|^2/2}\frac{\alpha_i^{n_i}}{\sqrt{n_i!}}$, $|\alpha_i|^2 = \bar{n}_i$ is the initial mean photon number for the mode. Now, by substituting $|\Psi(t)\rangle$ from Eq. (5) and $\hat{H}_I$ from Eq. (3) in the time-dependent Schrődinger equation $i\frac{\partial}{\partial t}|\Psi(t)\rangle = \hat{H}_I |\Psi(t)\rangle$, one may arrive at the following coupled differential equations for the atomic probability amplitudes

$$i\frac{d}{dt}\begin{pmatrix}G_1\\G_2\\G_3\end{pmatrix} = \begin{pmatrix}\bar{\alpha}_1 & v_1 e^{i\delta_1 t} & 0\\ v_1 e^{-i\delta_1 t} & \bar{\alpha}_2 & v_2 e^{-i\delta_2 t}\\ 0 & v_2 e^{i\delta_2 t} & \bar{\alpha}_3\end{pmatrix}\begin{pmatrix}G_1\\G_2\\G_3\end{pmatrix}. \tag{7}$$

Where,
$$\bar{\alpha}_1 = \chi_1(n_1)(n_1 - 1) + \chi_2(n_2)[(n_2 - 1),$$
$$\bar{\alpha}_2 = \chi_1(n_1)(n_1 + 1) + \chi_2(n_2)(n_2 - 1),$$
$$\bar{\alpha}_3 = \chi_1(n_1)(n_1 + 1) + \chi_2(n_2)(n_2 + 1),$$
$$v_1 = \frac{\lambda_1}{2}\sqrt{n_1 + 1}, \quad v_2 = \frac{\lambda_2}{2}\sqrt{n_2 + 1}. \tag{8}$$

The solution of Eqs. (7) are given as follows
$$G_1 = \sum_{j=1}^{3} B_j e^{i\xi_j t},$$
$$G_2 = -\frac{1}{v_1}\sum_{j=1}^{3} B_j(\bar{\alpha}_1 + \xi_j)e^{i(\xi_j - \delta_1)t},$$
$$G_3 = \frac{1}{v_1 v_2}\sum_{j=1}^{3} B_j[(\xi_j + \bar{\alpha}_2 - \delta_1)(\bar{\alpha}_2 + \xi_j) - v_1^2]e^{i(\xi_j - \delta_1 + \delta_2)t}, \tag{9}$$

By applying these initial conditions for atom and field and using (9), the $B_j$ coefficients read as

$$B_j = \frac{[(\Gamma_3 + \xi_k + \xi_l)\bar{\alpha}_1 + \xi_k \xi_l - \Gamma_4]q_{n_1}q_{n_2}}{\xi_{jk}\,\xi_{jl}}, \quad j \neq k = 1, 2, \tag{10}$$

where $\xi_{jk} = \xi_j - \xi_k$, $\xi_j$, $j = 1, 2$, are the roots of the following third-order algebraic equation

$$\xi^3 + h_1\xi^2 + h_2\xi + h_3 = 0, \qquad (11)$$

where

$$h_1 = \Gamma_1 + \Gamma_3 + \bar{\alpha}_3, \quad h_2 = \Gamma_1\Gamma_3 + \Gamma_4 + \bar{\alpha}_3\Gamma_3 - V_2^2,$$
$$h_3 = \Gamma_1\Gamma_4 + \bar{\alpha}_3\Gamma_4 - \bar{\alpha}_1 V_2^2, \quad \Gamma_1 = \delta_2 - \delta_1,$$
$$\Gamma_2 = \bar{\alpha}_2 - \delta_1, \quad \Gamma_3 = \delta_1 + \Gamma_2, \quad \Gamma_4 = \bar{\alpha}_1\Gamma_2 - V_1^2. \qquad (12)$$

The three roots of third-order Eq. (11) are given in the following form [38]

$$\xi_m = -\frac{1}{3}h_1 + \frac{2}{3}\sqrt{h_1^2 - 3h_2}\cos(\Phi + \frac{2}{3}(m-1)\pi), \quad m = 1,2,3,$$
$$\Phi = \frac{1}{3}\arccos[\frac{9h_1 h_2 - 2h_1^3 - 27h_3}{2(h_1^2 - 3h_2)^{2/3}}]. \qquad (13)$$

At any time $t > 0$ the reduced density matrix of the atom describing the system is given by:

$$\hat{\varrho}(t) = \begin{pmatrix} \varrho_{11}(t) & \varrho_{12}(t) & \varrho_{13}(t) \\ \varrho_{21}(t) & \varrho_{22}(t) & \varrho_{23}(t) \\ \varrho_{31}(t) & \varrho_{32}(t) & \varrho_{33}(t) \end{pmatrix}, \qquad (14)$$

where

$$\varrho_{11}(t) = \sum_{n_1,n_2=0}^{\infty} G_1(n_1, n_2, t)G_1^*(n_1, n_2, t),$$
$$\varrho_{22}(t) = \sum_{n_1,n_2=0}^{\infty} G_2(n_1 + 1, n_2, t)G_2^*(n_1 + 1, n_2, t),$$
$$\varrho_{33}(t) = \sum_{n_1,n_2=0}^{\infty} G_3(n_1 + 1, n_2 + 1, t)G_3^*(n_1 + 1, n_2 + 1, t), \ldots,$$
$$\varrho_{il}(t) = \varrho_{li}^*(t). \qquad (15)$$

In the next sections, for simplisity, we consider the constants $\lambda_i = \lambda$, have been taken to be real and the interaction time is the scaled time $\tau = \lambda t$.

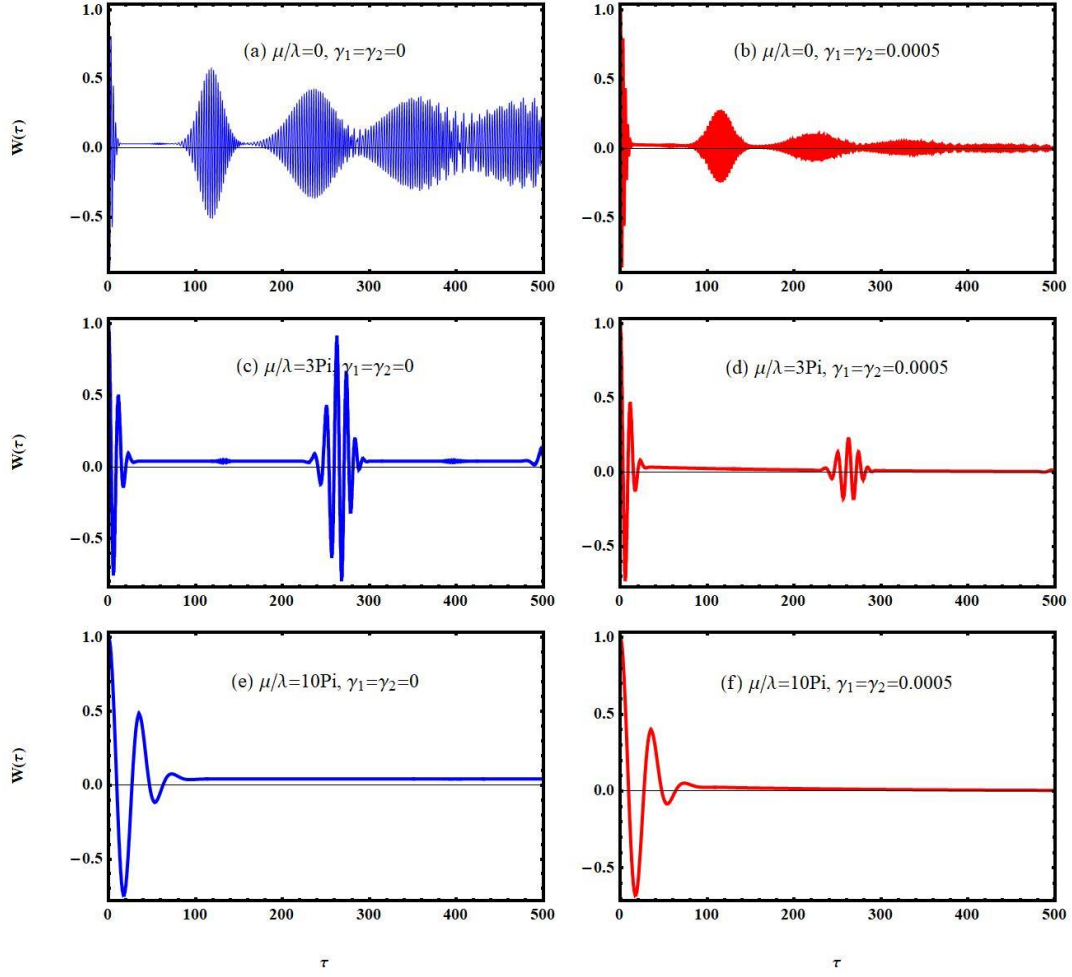

**Fig. 2** Evolution of the atomic population inversion $W(\tau)$ for of a three-level $\Xi$-type atom interacting with a two-mode coherent field for the parameters $\bar{n}_1 = \bar{n}_2 = 10$, $\Delta_1 = \Delta_2 = 0$, $\chi_1 = \chi_2 = 0$ and for: (a) $\mu/\lambda = 0$, $\gamma_1 = \gamma_2 = 0$, (b) $\mu/\lambda = 0$, $\gamma_1 = \gamma_2 = 0.0005$, (c) $\mu/\lambda = 3Pi$, $\gamma_1 = \gamma_2 = 0$, (d) $\mu/\lambda = 3Pi$, $\gamma_1 = \gamma_2 = 0.0005$, (e) $\mu/\lambda = 10Pi$, $\gamma_1 = \gamma_2 = 0$, (f) $\mu/\lambda = 10Pi$, $\gamma_1 = \gamma_2 = 0.0005$.

# 3 Atomic Population Inversion

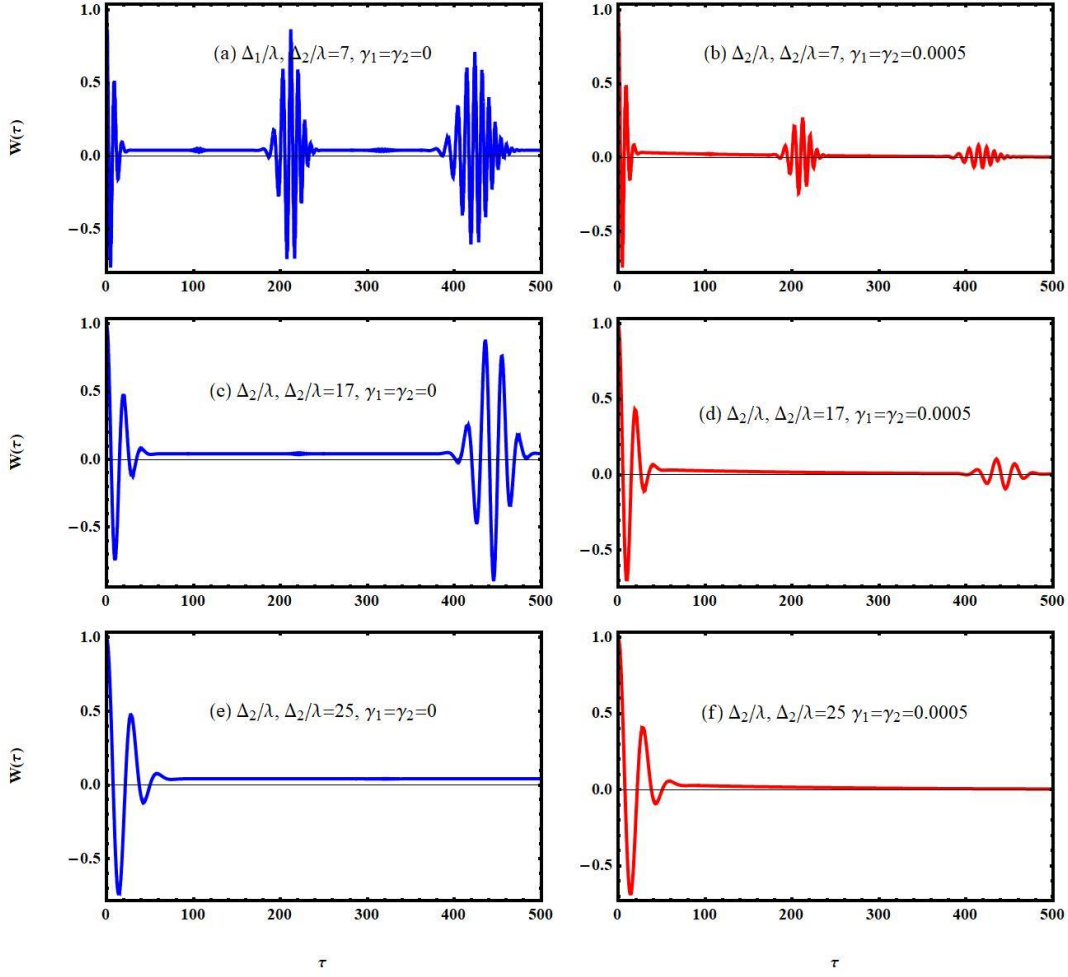

**Fig. 3** Evolution of the atomic population inversion $W(\tau)$ for of a three-level $\Xi$-type atom interacting with a two-mode coherent field for the parameters $\bar{n}_1 = \bar{n}_2 = 10$, $\mu/\lambda = 0, \chi_1 = \chi_2 = 0$ and for: (a) $\Delta_1/\lambda = \Delta_2/\lambda = 7$, $\gamma_1 = \gamma_2 = 0$, (b) $\Delta_1/\lambda = \Delta_2/\lambda = 7, \gamma_1 = \gamma_2 = 0.0005$, (c) $\Delta_1/\lambda = \Delta_2/\lambda = 17$, $\gamma_1 = \gamma_2 = 0$, (d) $\Delta_1/\lambda = \Delta_2/\lambda = 17$, $\gamma_1 = \gamma_2 = 0.0005$, (e) $\Delta_1/\lambda = \Delta_2/\lambda = 25$, $\gamma_1 = \gamma_2 = 0$, (f) $\Delta_1/\lambda = \Delta_2/\lambda = 25$, $\gamma_1 = \gamma_2 = 0.0005$.

In fact, we can get information about the behavior of the atom-field interaction through the collapse and revival phenomenon. So, we shall study the dynamics of an important quantity, namely atomic population inversion. The atomic inversion is defined as the difference between the exited state $|1\rangle$ and the ground state $|3\rangle$ which may be written as follows [3]

$$W(t) = \varrho_{11}(t) - \varrho_{33}(t). \tag{16}$$

Now, we shall study the behavior of the atomic population inversion in the time-dependent case for $\bar{n}_1 = \bar{n}_2 = 10$. This will be done on the basis of the previous calculations. We examine the influence of the time-dependent coupling parameter, detuning parameters, Kerr-medium on the behavior of the atomic population inversion in the absence or presence of the photon assisted atomic phase damping parameter. The temporal evolution atomic population inversion has given in Figs. 2-4 versus scaled time $\tau = \lambda t$. The left plot for $\gamma_1 = \gamma_1 = 0$ and the right plot for $\gamma_1 = \gamma_1 = 0.0005$. In Fig. 2(a) and Fig. 2(b), we have considered the time-dependent coupling parameter $\mu/\lambda = 0$ in the absence of the detuning parameters and Kerr-medium ($\Delta_1/\lambda = \Delta_2/\lambda = \chi_1 = \chi_2 = 0$). The behavior of the atomic population inversion in Fig. 2(a) exhibits the collapse and revival phenomena. The number of oscillations in Fig. 2(b) is less than that in Fig. 2(a). Also, in Fig. 2(b), the effect of the photon assisted atomic phase damping parameter leads to decreasing the amplitude of oscillations as time develops and the mean value of oscillations become zero in the time evolution process. In Fig. 2(c) and Fig. 2(d) in which the value of the time-dependent coupling parameter $\mu/\lambda = 3Pi$, the behavior of the atomic population inversion in Fig. 2(c) is changed compared with Fig. 2(a). The collapse intervals is elongated. In Fig. 2(e), when $\mu/\lambda = 10Pi$, we observe that the behavior of the atomic population inversion has started only with a short period of revivals followed by a long time-interval of collapse compared with the previous cases. This means that we can consider the time-dependent coupling as a quantum control parameter. The effect of the detunning parameters on the atomic population inversion in the absence or presence of the photon assisted atomic phase damping parameter and in the absence of both of the time-dependent coupling parameter and Kerr-medium ($\mu/\lambda = \chi_1 = \chi_2 = 0$) appeared in Fig. 3. In Fig. 3(a), when $\Delta_1/\lambda = \Delta_2/\lambda = 7$, $\gamma_1 = \gamma_1 = 0$, we have along intervals of collapses compared with that in Fig. 2(a). Also, in Fig. 3(b), the effect of the photon assisted atomic phase damping parameter leads to decreasing the amplitude of oscillations as time develops and the mean value of oscillations become zero in the time evolution process. By the increase of the value of the detuning parameter the collapses interval is elongated so, we can con consider the detuning parameter as a quantum control parameter (see Figs 3(c-f)). The behavior of the atomic population inversion in Fig. 3(e) and Fig. 3(f) is similar to that in Fig. 2(e) and Fig. 2(f) ($\Delta_1/\lambda = \Delta_3/\lambda = 25$). To discuss the influence of Kerr-medium on the atomic population inversion in the absence or presence of the photon assisted atomic phase damping parameter also in the absence of both of the time-dependent coupling parameter and detuning parameters ($\mu/\lambda = \Delta_1/\lambda = \Delta_2/\lambda = 0$), we have plotted Fig. 4. For a small value of Kerr-medium parameter ($\chi_1 = \chi_2 = 0.01$), the behavior of $W(\tau)$ in Fig. 4(a) is changed compared to the behavior of $W(\tau)$ in Fig. 2(a) and Fig. 3(a), the amplitude of oscillations is decreased. By the increase of the value of Kerr-medium, the behavior of $W(\tau)$ changes. The mean value of oscillations is shifted upward. For a great value of Kerr-medium the behavior of the atomic population inversion is completely changed compared with the previous cases. We have a greatest negative mean value of oscillations and the maximum value of fluctuations approaches to one (see Fig. 4(e)). This means that the energy increases in the atomic system. The photon assisted atomic phase damping parameter leads to destroy the amplitude of oscillations as time develops (see Fig. 4(b, d, f)).

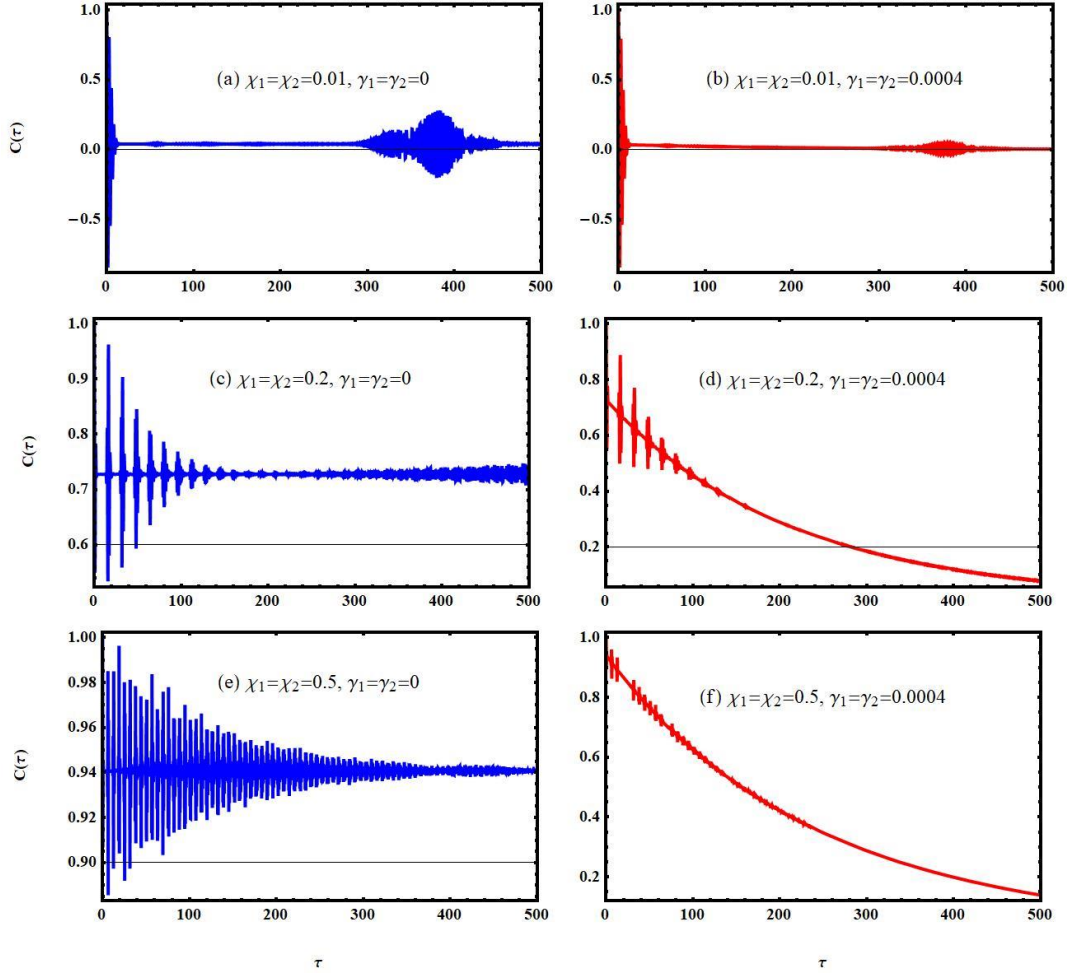

**Fig. 4** Evolution of the atomic population inversion $W(\tau)$ for of a three-level $\Xi$-type atom interacting with a two-mode coherent field for the parameters $\bar{n}_1 = \bar{n}_2 = 10$, $\mu/\lambda = 0$, $\Delta_1 = \Delta_2 = 0$ and for: (a) $\chi_1 = \chi_2 = 0.01$, $\gamma_1 = \gamma_2 = 0$, (b) $\chi_1 = \chi_2 = 0.01, \gamma_1 = \gamma_2 = 0.0004$, (c) $\chi_1 = \chi_2 = 0.2$, $\gamma_1 = \gamma_2 = 0$, (d) $\chi_1 = \chi_2 = 0.2$, $\gamma_1 = \gamma_2 = 0.0004$, (e) $\chi_1 = \chi_2 = 0.5$, $\gamma_1 = \gamma_2 = 0$, (f) $\chi_1 = \chi_2 = 0.5, \gamma_1 = \gamma_2 = 0.0004$.

## 4   Concurrence

The concurrence is presented by Wootters and Hill [39, 40] as a proper measure of the entanglement of any state of two qubits, pure or mixed. For a pure state $|\Psi(t)\rangle$ on $(K \times L)$-dimensional Hilbert space $M = M_K \otimes M_L$. The concurrence can be defined as follows [10, 41]

$$C(t) = \sqrt{2[|\langle\Psi(t)|\Psi(t)\rangle|^2 - Tr(\varrho_L^2(t))]},$$

where $\varrho_L(t) = Tr_K(|\Psi(t)\rangle\langle\Psi(t)|)$ is the reduced density operator of the subsystem with dimension $L$ and $Tr_K$ is the partial trace over $M_K$. It is remarkable to mention that, the concurrence fluctuates between $\sqrt{2(L-1)}$ for maximally entangled state and 0 for separable state. Herein we calculate the concurrence to get the degree of entanglement (DEM) between the atom and the field. Using Eq. (14), we can rewrite concurrence in the following form [17]

$$C(t) = \sqrt{2\sum_{i,j=1,2,\dots 9}^{i\neq j}[\varrho_{ii}(t)\varrho_{jj}(t) - \varrho_{ij}(t)\varrho_{ji}(t)]}.$$

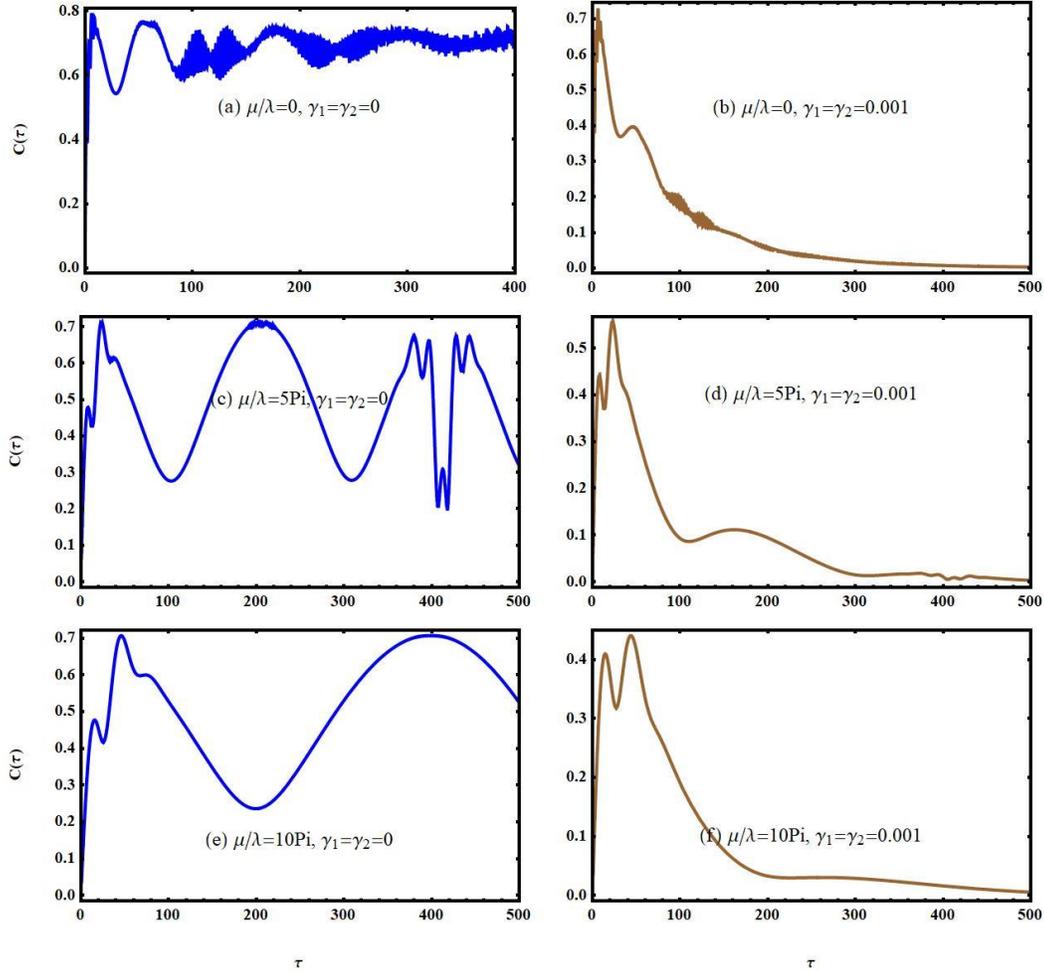

**Fig. 5** Evolution of the concurrence $C(\tau)$ for of a three-level $\Xi$-type atom interacting with a two-mode coherent field for the parameters $\bar{n}_1 = \bar{n}_2 = 10$, $\Delta_1 = \Delta_2 = 0$, $\chi_1 = \chi_2 = 0$ and for: (a) $\mu/\lambda = 0$, $\gamma_1 = \gamma_2 = 0$, (b) $\mu/\lambda = 0, \gamma_1 = \gamma_2 = 0.001$, (c) $\mu/\lambda = 5Pi$, $\gamma_1 = \gamma_2 = 0$, (d) $\mu/\lambda = 5Pi$, $\gamma_1 = \gamma_2 = 0.001$, (e) $\mu/\lambda = 10Pi$, $\gamma_1 = \gamma_2 = 0$, (f) $\mu/\lambda = 10Pi$, $\gamma_1 = \gamma_2 = 0.001$.

Now, we are going to study the evolution of the concurrence $C(\tau)$ versus the scaled time $\tau = \lambda t$ for the same parameters that we used in Figs 2-4. An illustration of the time evolution of the concurrence for $\gamma_1 = \gamma_1 = 0$ (left plot), $\gamma_1 = \gamma_1 = 0.001$ (right plot) and for $\bar{n}_1 = \bar{n}_2 = 10$ is shown in Figs. 5-7. In Fig. 5(a), when all parameters are zero, we not that $C(\tau)$ starts from zero, then it followed by a sequence of fluctuations in the oscillation. This means that this system begins by disentangled state then it develops to a mixed state ($t > 0$). It is clear that, in Fig. 5(c) and Fig. 5(e), the time-dependent coupling parameter plays a dramatic role in the degree of entanglement, the maximum value of $C(\tau)$ is decreased and the periodic behavior is appeared. Also, the time interval of the period is elongated as the value of $\mu/\lambda$ increases. In Figs. 5(b, d, f), we observed that the photon assisted atomic phase damping parameter ($\gamma_1 = \gamma_1 = 0.001$) leads to deceases the degree of entanglement between the atom and the field and finally vanishes as the time develops (i.e. no entanglement).

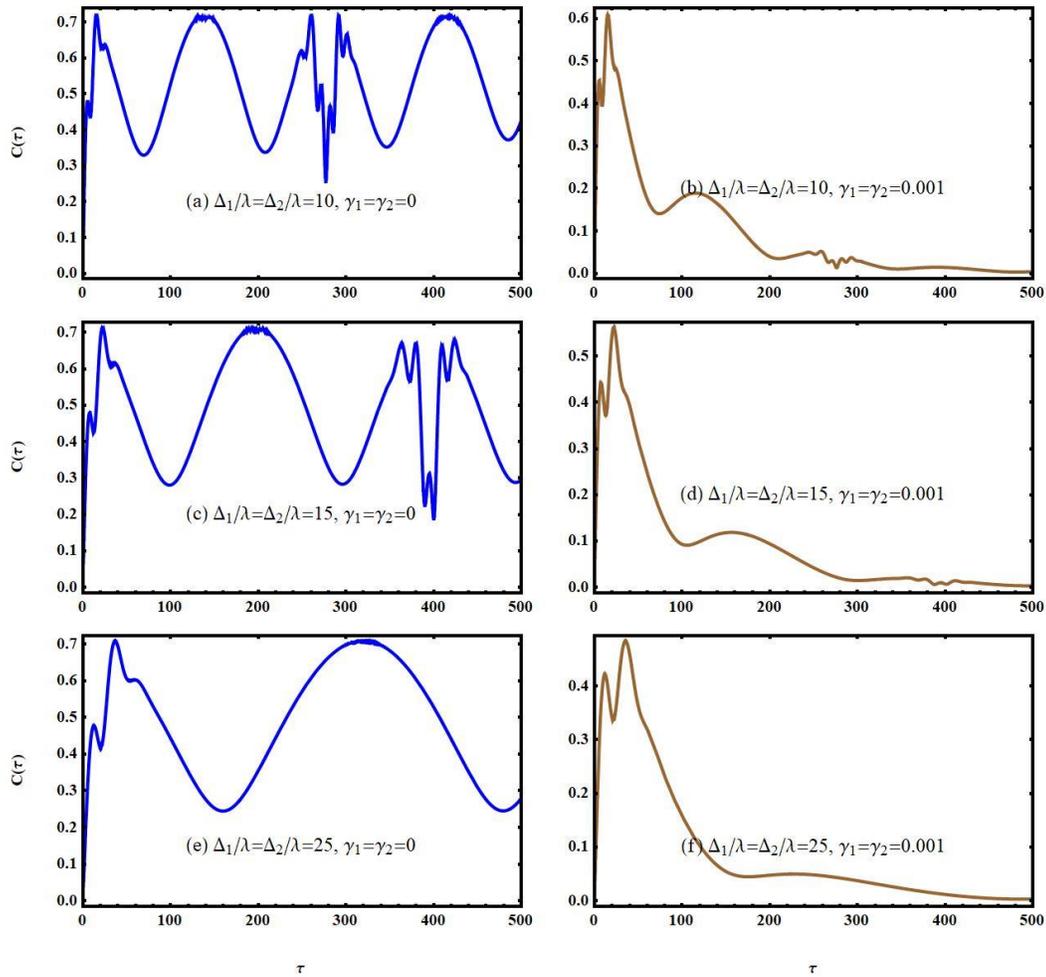

**Fig. 6** Evolution of the concurrence $C(\tau)$ for of a three-level $\Xi$-type atom interacting with a two-mode coherent field for the parameters $\bar{n}_1 = \bar{n}_2 = 10$, $\mu/\lambda = 0$, $\chi_1 = \chi_2 = 0$ and for: (a) $\Delta_1/\lambda = \Delta_2/\lambda = 10$, $\gamma_1 = \gamma_2 = 0$, (b) $\Delta_1/\lambda = \Delta_2/\lambda = 10$, $\gamma_1 = \gamma_2 = 0.001$, (c) $\Delta_1/\lambda = \Delta_2/\lambda = 15$, $\gamma_1 = \gamma_2 = 0$ , (d) $\Delta_1/\lambda = \Delta_2/\lambda = 15$, $\gamma_1 = \gamma_2 = 0.001$, (e) $\Delta_1/\lambda = \Delta_2/\lambda = 25$, $\gamma_1 = \gamma_2 = 0$, (f) $\Delta_1/\lambda = \Delta_2/\lambda = 25$, $\gamma_1 = \gamma_2 = 0.001$.

To explore the effect of the detuning parametes $\Delta_1/\lambda$, $\Delta_2/\lambda$, on $C(\tau)$ in the absence or presence of the photon assisted atomic phase damping parameter and in the absence of both of the time-dependent coupling parameter and Kerr-medium ($\mu/\lambda = \chi_1 = \chi_2 = 0$), we have plotted Fig. 6. In Fig 6(a), when $\Delta_1/\lambda = \Delta_2/\lambda = 10$, $\gamma_1 = \gamma_1 = 0$, we observed that the maximum value of $C(\tau)$ is decreased compared with Fig. 5(a). Also, the periodic behavior is appeared in Fig. 6(a). As the detuning parameter increases, the time interval of the period is elongated (see Fig. 6(c) and Fig. 6(e)). The effect of detuning parameter in the presence of the photon assisted atomic phase damping parameter leads to decreases the degree of entanglement between the atom and the field as the time develops (see Figs. 6(b, d ,f)). To visualize the influence of Kerr-medium on the concurrence $C(\tau)$ in the absence or presence of the photon assisted atomic phase damping parameter and in the absence of both of the time-dependent coupling parameter and detuning parameters ($\mu/\lambda = \Delta_1/\lambda = \Delta_2/\lambda = 0$), we have plotted Fig. 7. We notice that when $\chi_1/\lambda = \chi_2/\lambda = 0.01$, $\gamma_1 = \gamma_1 = 0$, the nonlinear interaction of the Kerr-medium with the field modes leads to increasing the maximum value of the concurrence with decreasing of the amplitude of oscillations (see Fig. 7(a)). By the increase of the nonlinear interaction of the Kerr-medium with field modes, the maximum value of the concurrence decreases and then the degree of entanglement between the atom and the filed decreases (see Fig. 7(c) and Fig. 7(e)). Also, we observed that when $\chi_1/\lambda = \chi_2/\lambda = 0.5$, $\gamma_1 = \gamma_1 = 0$, many oscillations have appeared. The effect of Kerr-medium in the presence of the photon assisted atomic phase damping parameter leads to decreases the degree of entanglement between the atom and the field in the time evolution process (see Figs. 7(b, d ,f)).

## 5    Conclusion

In summary, we have studied a three-level $\Xi$-type atom interacting with a two-mode field. The photon assisted atomic phase damping parameter, Kerr-medium and the detuning parameter are taken into account. Also, the coupling parameter is modulated to be time-dependent. Under an approximation similar to that of the Rotating-Wave Approximation (RWA), the exact expression of atom-field wave function is obtained. After obtaining the exact analytical form of the state vector of the whole system, the influence of the photon assisted atomic phase damping parameter, the time-dependent coupling parameter, detuning parameter and Kerr nonlinearity on the atomic population inversion and the concurrence of the system have been studied. The investigations have shown that the atomic population inversion has the quantum collapse-revival behavior. Also, according to the previous discussion, the time-dependent coupling parameter and detuning parameter can be considered as a quantum control parameters. The concurrence of a three-level atomic system has been introduced and its time evolution has been studied which provides the ability to explore the degree of entanglement of the available systems in the absence or presence the photon assisted atomic phase damping parameter. Finally, we can deduce that the presence of the time-dependent coupling parameter, detuning parameter, Kerr nonlinearity and the photon assisted atomic phase damping parameter leads to a noticeable effects in the quantum entanglement of the considered systems. It is interesting to mention that one can study this system when both the field and the atom are initially prepared in other states.

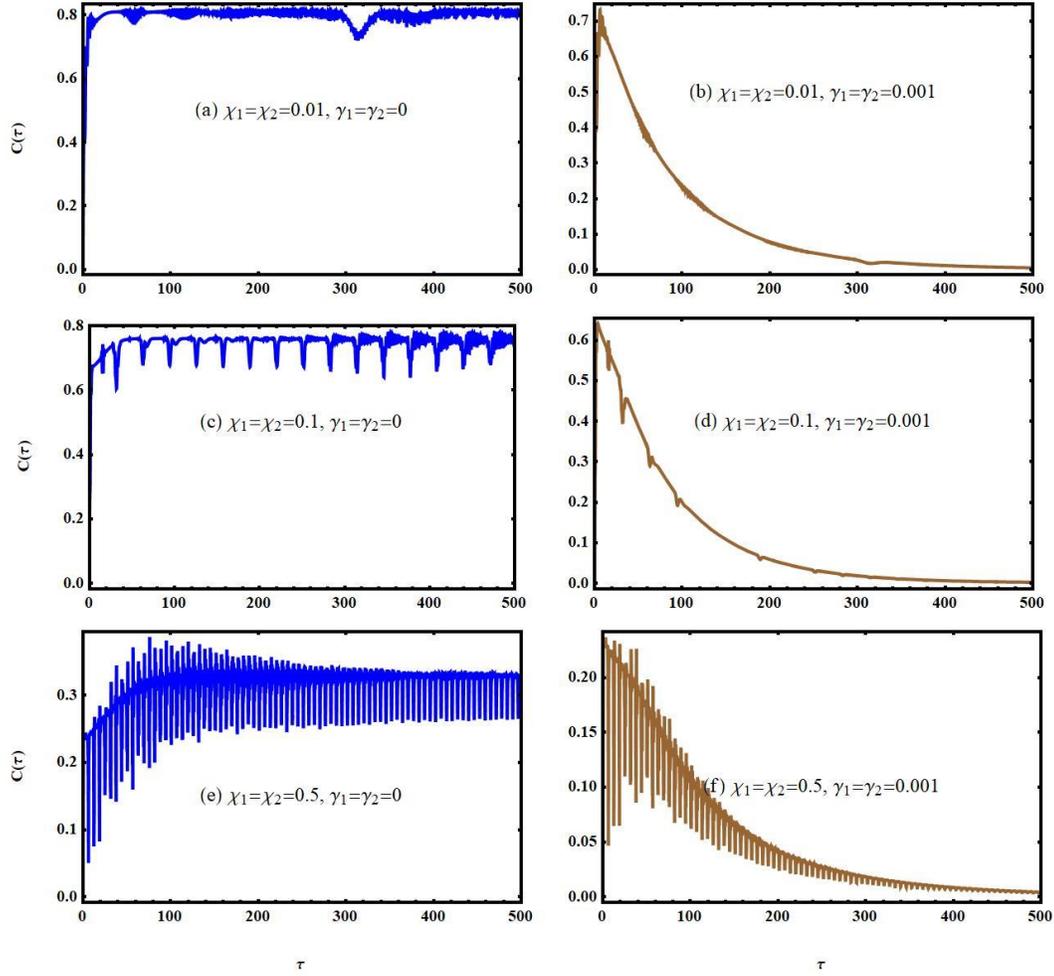

**Fig. 7** Evolution of the concurrence $C(\tau)$ for of a three-level $\Xi$-type atom interacting with a two-mode coherent field for the parameters $\bar{n}_1 = \bar{n}_2 = 10$, $\mu/\lambda = 0$, $\Delta_1 = \Delta_2 = 0$ and for: (a) $\chi_1 = \chi_2 = 0.01, \gamma_1 = \gamma_2 = 0$, (b) $\chi_1 = \chi_2 = 0.01$, $\gamma_1 = \gamma_2 = 0.001$, (c) $\chi_1 = \chi_2 = 0.1$, $\gamma_1 = \gamma_2 = 0$, (d) $\chi_1 = \chi_2 = 0.1$, $\gamma_1 = \gamma_2 = 0.001$, (e) $\chi_1 = \chi_2 = 0.5, \gamma_1 = \gamma_2 = 0$, (f) $\chi_1 = \chi_2 = 0.5$, $\gamma_1 = \gamma_2 = 0.001$.